\documentclass[longauth]{aa}  

\usepackage{graphicx}
\usepackage{natbib}
\usepackage{xcolor} 
\usepackage{hyperref}
\newcommand{\gaia}{Gaia\xspace}

\bibpunct{(}{)}{;}{a}{}{,} 
\usepackage[varg]{txfonts}
\urldef\rvperformanceurl\url{https://www.cosmos.esa.int/web/gaia/science-performance#spectroscopic%20performance}

\providecommand{\kms}{\ensuremath{\rm \,km\,s^{-1}}\xspace}

\begin{document}

   \title{\textit{Gaia}~Data Release 3: Hot-star radial velocities}

   \subtitle{}

   \author{
          R.~Blomme\inst{\ref{inst:01}}\thanks{\email{Ronny.Blomme@oma.be}}
          \and Y.~Fr\'{e}mat\inst{\ref{inst:01}}
          \and P.~Sartoretti\inst{\ref{inst:02}}
          \and A.~Guerrier\inst{\ref{inst:03}}
      \and P.~Panuzzo\inst{\ref{inst:02}}
      \and D.~Katz\inst{\ref{inst:02}}
          \and G. M.~Seabroke\inst{\ref{inst:04}}
      \and F.~Th\'{e}venin\inst{\ref{inst:05}}
      \and M.~Cropper\inst{\ref{inst:04}}
      \and K.~Benson\inst{\ref{inst:04}}
      \and Y.~Damerdji\inst{\ref{inst:06},\ref{inst:07}}
      \and R.~Haigron\inst{\ref{inst:02}}
      \and O.~Marchal\inst{\ref{inst:08}}
      \and M.~Smith\inst{\ref{inst:04}}
      \and S.~Baker\inst{\ref{inst:04}}
      \and L.~Chemin\inst{\ref{inst:09}}
      \and M.~David\inst{\ref{inst:10}}
      \and C.~Dolding\inst{\ref{inst:04}}
      \and E.~Gosset\inst{\ref{inst:07},\ref{inst:11}}
      \and K.~Jan{\ss}en\inst{\ref{inst:12}}
      \and G.~Jasniewicz\inst{\ref{inst:13}}
      \and A.~Lobel\inst{\ref{inst:01}}
      \and G.~Plum\inst{\ref{inst:02}}
      \and N.~Samaras\inst{\ref{inst:01},\ref{inst:14}}
      \and O.~Snaith\inst{\ref{inst:02}}
      \and C.~Soubiran\inst{\ref{inst:15}}
      \and O.~Vanel\inst{\ref{inst:02}}
      \and T.~Zwitter\inst{\ref{inst:16}}
      \and N.~Brouillet\inst{\ref{inst:15}}
      \and E.~Caffau\inst{\ref{inst:02}}
      \and F.~Crifo\inst{\ref{inst:02}} 
      \and C.~Fabre\inst{\ref{inst:17},\ref{inst:03}}
      \and F.~Frakgoudi\inst{\ref{inst:02}}
      \and H.E.~Huckle\inst{\ref{inst:04}}
      \and A.~Jean-Antoine Piccolo\inst{\ref{inst:03}}
      \and Y.~Lasne\inst{\ref{inst:03}}
      \and N.~Leclerc\inst{\ref{inst:02}}
      \and A.~Mastrobuono-Battisti\inst{\ref{inst:02},\ref{inst:18}}
      \and F.~Royer\inst{\ref{inst:02}}
      \and Y.~Viala\inst{\ref{inst:02}}
      \and J.~Zorec\inst{\ref{inst:19}}
    }

   \institute{
        Royal Observatory of Belgium, Ringlaan 3, B-1180 Brussel, Belgium
        \label{inst:01}
        \and  GEPI, Observatoire de Paris, Universit\'{e} PSL, CNRS, 5 Place Jules Janssen, F-92190 Meudon, France
        \label{inst:02}
        \and CNES Centre Spatial de Toulouse, 18 avenue Edouard Belin, F-31401 Toulouse Cedex 9, France                                                                                                        
        \label{inst:03}
        \and
        Mullard Space Science Laboratory, University College London, Holmbury St Mary, Dorking, Surrey, RH5 6NT, United Kingdom\label{inst:04}
        \and Universit\'{e} C\^{o}te d’Azur, Observatoire de la C\^{o}te d’Azur, CNRS, Laboratoire Lagrange, Boulevard de l’Observatoire, CS 34229, 06304 Nice, France
        \label{inst:05}
        \and CRAAG - Centre de Recherche en Astronomie, Astrophysique et G\'{e}ophysique, Route de l'Observatoire, Bp 63 Bouzareah, DZ-16340, Alger, Alg\'{e}rie                                            
        \label{inst:06}
        \and Institut d'Astrophysique et de G\'{e}ophysique, Universit\'{e} de Li\`{e}ge, 19c, All\'{e}e du 6 Ao\^{u}t, B-4000 Li\`{e}ge, Belgium
        \label{inst:07}
        \and Observatoire astronomique de Strasbourg, Universit\'{e} de Strasbourg, CNRS, 11 rue de l'Universit\'{e}, F-67000 Strasbourg, France
        \label{inst:08}
        \and Centro de Astronom\'ia, Universidad de Antofagasta, Avda. U. de Antofagasta, 02800 Antofagasta, Chile
        \label{inst:09}
        \and Universiteit Antwerpen, Onderzoeksgroep Toegepaste Wiskunde, Middelheimlaan 1, B-2020 Antwerpen, Belgium                                                                                         
        \label{inst:10}
        \and F.R.S.-FNRS, Rue d'Egmont 5, B-1000 Brussels, Belgium                                                                                                                                           
        \label{inst:11}
        \and Leibniz Institute for Astrophysics Potsdam (AIP), An der Sternwarte 16, D-14482 Potsdam, Germany
        \label{inst:12}
        \and Laboratoire Univers et Particules de Montpellier, Universit\'{e} Montpellier, CNRS, Place Eug\`{e}ne Bataillon, CC72, F-34095 Montpellier Cedex 05, France                                      
        \label{inst:13}
        \and Charles University, Faculty of Mathematics and Physics, Astronomical Institute of Charles University, V Hole\v{s}ovi\v{c}k\'{a}ch 2, 180 00 Prague, Czech Republic
        \label{inst:14}
        \and Laboratoire d'astrophysique de Bordeaux, Universit\'{e} de Bordeaux, CNRS, B18N, all{\'e}e Geoffroy Saint-Hilaire, F-33615 Pessac, France\label{inst:15}
        \and Faculty of Mathematics and Physics, University of Ljubljana, Jadranska ulica 19, SLO-1000 Ljubljana, Slovenia                                                                               
        \label{inst:16}
        \and ATOS for CNES Centre Spatial de Toulouse, 18 avenue Edouard Belin, 31401 Toulouse Cedex 9, France
        \label{inst:17}
        \and Department of Astronomy and Theoretical Physics, Lund Observatory, Box 43, SE--221 00, Lund, Sweden
        \label{inst:18}
        \and Sorbonne Université CNRS, UMR 7095, Institut d’Astrophysique
        de Paris, 75014 Paris, France\label{inst:19}
             }

   \date{Received <date>; accepted <date>}

  
  \abstract
  {The second \textit{Gaia} data release, DR2, contained radial velocities of stars
        with effective temperatures
        up to $T_\mathrm{eff} = 6900$~K. The third data release, \textit{Gaia} DR3, extends
        this up to $T_\mathrm{eff} = 14\,500$~K.
   }
   {We derive the radial velocities for hot stars 
        (i.e. in the $T_\mathrm{eff} = 6900 - 14\,500$~K range) from data
        obtained with the Radial Velocity Spectrometer (RVS) on board \textit{Gaia}.
   }
   {The radial velocities were determined
        by the standard technique of measuring the Doppler shift of a template
        spectrum that was compared to the observed spectrum.
        The RVS wavelength range is very limited. The proximity
        to and systematic blueward offset of the 
        calcium infrared triplet to the hydrogen Paschen lines 
        in hot stars can result in a systematic
        offset in radial velocity. 
        For the hot stars, we developed a specific 
        code to improve the selection
        of the template spectrum, thereby avoiding this systematic offset.
   }
   {With the improved code, and with the correction we propose to
        the DR3 archive radial velocities, we obtain values that agree 
        with reference values 
        to within 3 \kms (in median). Because of the required
        S/N for applying the improved code, the hot star
        radial velocities in DR3 are mostly limited to stars with a magnitude
        in the RVS wavelength band $\le 12$~mag.
   }
   {}

   \maketitle
%

\section{Introduction}

The \textit{Gaia} satellite \citep{Gaia} was launched in December 2013.
By continuously scanning the sky, it is collecting astrometric, photometric, and spectroscopic
information on a large number of stars.
A number of data releases have made part of this wealth of information public
(DR1 by  \citealt{Gaia-DR1}; DR2 by \citealt{Gaia-DR2}; EDR3 by \citealt{Gaia-EDR3}).
Now the third data release \citep[\textit{Gaia}~DR3, ][]{Gaia-DR3} is available, which not only updates the 
previous releases, but also adds substantially new information, such as 
BP/RP spectra \citep{BPRPspectra},
astrophysical parameters \citep{DR3-DPACP-157}, 
and variability classification \citep{Eyer+22}.
This new information also includes the radial velocities
for almost 34 million stars \citep{Katz+22}, and Radial Velocity Spectrometer (RVS) spectra for almost one million stars
\citep{Seabroke+22}.

The \textit{Gaia} data are processed by the Data Processing and Analysis Consortium
\citep[DPAC; see][for details]{Gaia-DR1}.
Within DPAC, the Coordination Unit 6 (CU6) is responsible for 
handling the RVS spectra. Its main tasks are the reduction of the RVS
data for use by other Coordination Units and for publication in the data releases, as well as the determination of the radial
velocity from these spectra. A secondary task is measuring the broadening
of the spectral lines. This broadening is mainly due to the (projected) rotational velocity
of the star, but may also include other effects, such as macroturbulence
\citep{Fremat+22}. For
this reason, we use the term `broadening velocity' rather than 
rotational velocity. Other uses of the spectra, such as astrophysical
parameter determination, are within the remit of other DPAC Coordination Units.

This paper is one of four presenting the radial velocity results in DR3. 
\cite{Katz+22} present the DR3 radial velocity content in general, 
\cite{Gosset+22} describe the processing and performance for SB1 binary stars in detail,
and \cite{Damerdji+22} does the same for SB2 binaries. The present paper focusses on the radial velocity
performances of hot stars.
For a scientific application in which the radial velocities play an important
role, we refer to \citet{Drimmel+22}.

Data release 3 is the first data release to contain radial velocities for stars with effective temperatures
above 6900 K, extending the range up to 14\,500~K. 
The previous data release DR2 did not include these, as the quality of the results
could not be guaranteed. The present paper describes how these hot-star radial velocities were
derived, and the caveats that have to be taken into account in their use.
For stars even hotter than 14\,500~K, there remain
challenges that could
not be solved in DR3 and will therefore be handled in the next release, DR4.

\begin{figure*}
        \begin{center}
                \resizebox{\hsize}{!}{\includegraphics[bb=0 9 926 210,clip]{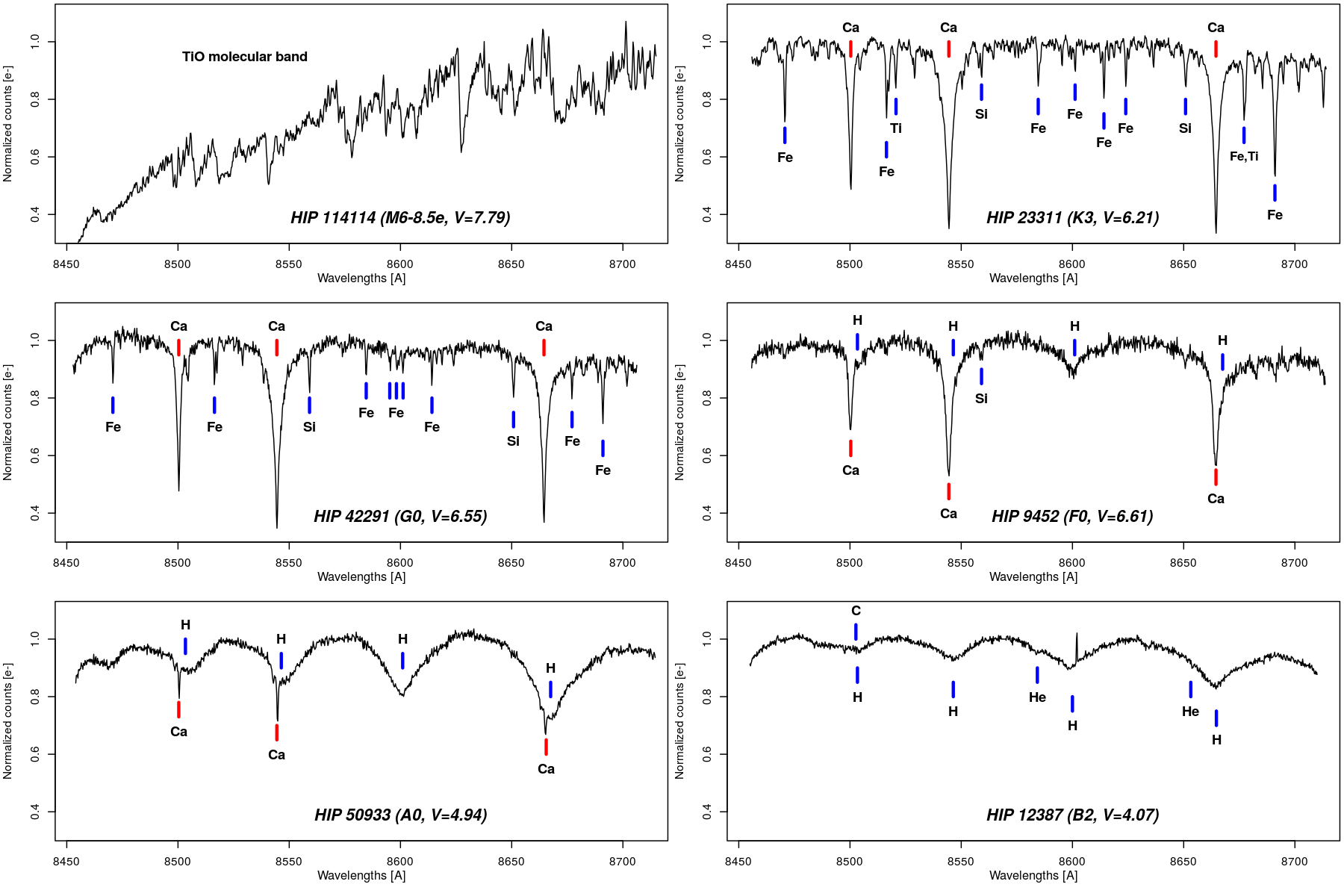}}
        \end{center}
        \caption{Two examples of RVS spectra of hot stars (\protect\url{https://www.cosmos.esa.int/web/gaia/iow_20141124}).
                Both stars have Paschen hydrogen lines, and the cooler star
                (left panel)
                also shows the calcium infrared triplet lines.
        A spectrum of the cooler star with a much higher S/N is available
in the DR3 archive (\object{Gaia DR3 1053778957742409984}).}
        \label{figure hot stars}
\end{figure*}

Determining radial velocities for hot stars with the \textit{Gaia} RVS is quite challenging. The 
instrument is described in detail in \citet{Cropper+18}. It is
a medium-resolution spectrograph ($R \approx 11\,500$) covering the wavelength range 846 -- 870 nm.
This rather small wavelength range has been chosen to optimise radial velocity determination
for cool stars. Figure~\ref{figure hot stars} shows two examples of the RVS spectrum of a hot star.
The hotter star (right panel) is dominated by the hydrogen Paschen lines, with no other substantial spectral lines
that can be used for radial velocity determination. The somewhat cooler star
(left panel) has, in addition to the Paschen lines, also the lines from
the calcium infrared triplet. While these sharper calcium lines are useful for radial velocity 
measurements, their proximity to the Paschen lines complicates matters.

The focal plane of \textit{Gaia} contains four rows of three CCDs each that are dedicated to the
RVS instrument. During a single transit of a star across the focal plane, 
three spectra are therefore taken. 
The integration time for a single spectrum is $\sim 4.4$~s.
During the 34 months of observations
covered by DR3 each star has 22 of these transits on average. The information from these 22 transits can be combined to improve the signal-to-noise ratio (S/N) and thus the 
precision of the radial velocity determination. 

In Sect.~\ref{section method} we present the part of the pipeline
that determines the radial velocity, and discuss the resulting hot-star radial
velocities from
a preliminary run of the code.
Section~\ref{section improved method} presents the specific method we
implemented to improve the handling of the hot stars. In Sect.~\ref{section results}
we show the improvement in the resulting hot-star radial velocities.
Section~\ref{section conclusions} presents the conclusions and the caveats
for the user of the \textit{Gaia} DR3 hot-star radial velocities.

\begin{figure}
        \begin{center}
                \resizebox{\hsize}{!}{\includegraphics[bb=45 80 330 370,clip]{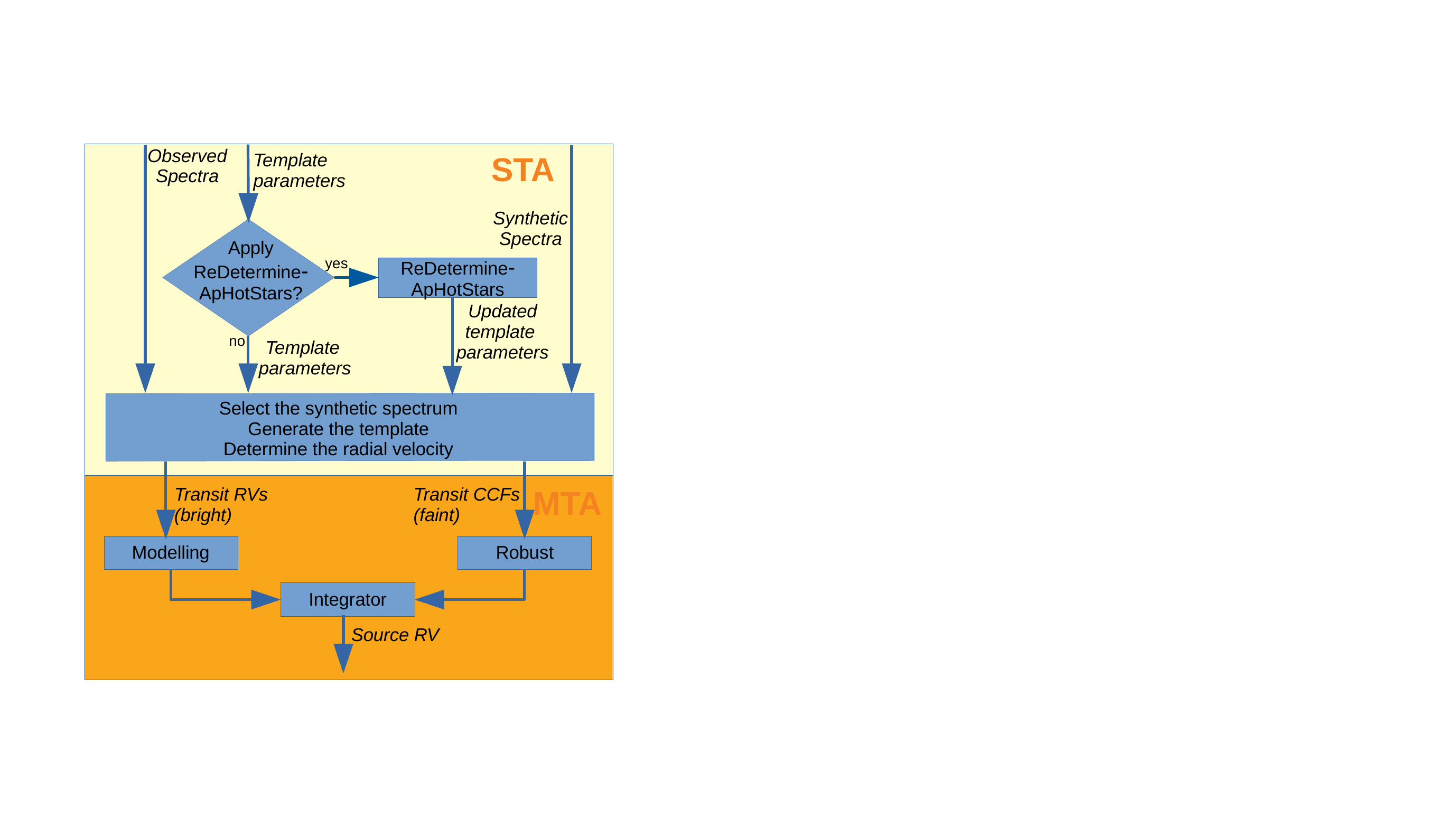}}
                \caption{Schematic overview of the STA and MTA parts of the pipeline.
                        The input to STA consists of the normalised observed spectra and the
                        template parameters. If ReDetermineApHotStars is applied 
                        (see Sect.~\ref{section improved method}), the 
                        template parameters are updated. The template parameters are then
                        used to select the corresponding synthetic spectrum; this is converted
                        into a template and used to derive the cross-correlation functions and
                        the radial velocity for each transit. 
                        The MTA part of the pipeline takes the time
                        series of transit radial velocities 
                        (for bright stars, i.e. \texttt{grvs\_mag} $\le 12$~mag),
                        or the time series of the transit cross-correlation functions
                        (for faint stars, i.e. \texttt{grvs\_mag} $> 12$~mag) to determine the stellar
                        radial velocity.
                }
                \label{figure STAMTA}
        \end{center}
\end{figure}

\begin{figure}[t]
        \begin{center}
                \resizebox{\hsize}{!}{\includegraphics[bb=46 34 300 185,clip]{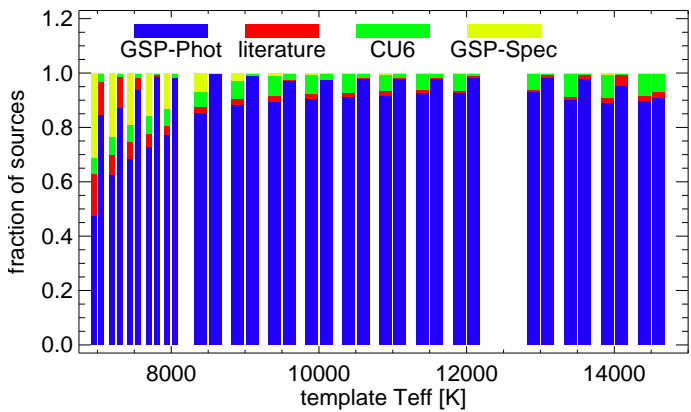}}
                \caption{Histogram showing the fraction of the stars
                        that obtained their template parameters from the various input
                        possibilities. The x-axis shows the template effective temperature.
                        Each histogram bar is split into two parts: the left-hand side of each bar shows 
                        the distribution for 
                    \texttt{grvs\_mag} $\le 12$~mag, and the right-hand side of each bar is
                    for \texttt{grvs\_mag} $>12$~mag.
                }
                \label{figure stats}
        \end{center}
\end{figure}

\begin{figure*}[t]
        \begin{center}
                \includegraphics[bb=25 10 530 160]{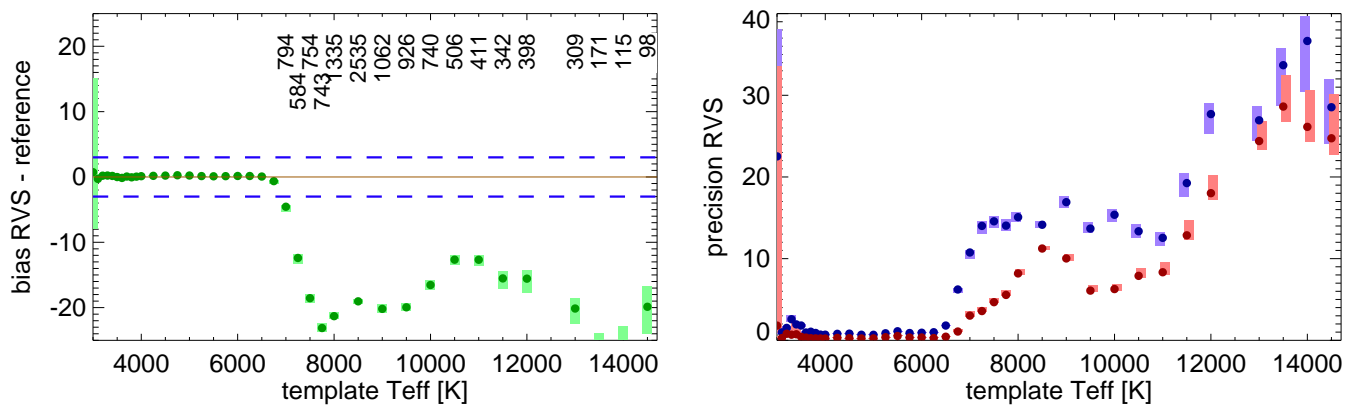}
                \caption{Preliminary radial velocities compared to their reference 
                        values as a function of the template effective temperature.
                        ReDetermineApHotStars was not applied to these data.
                        The \emph{left panel} shows the bias of the radial velocity of
                        the stars (RVS value minus reference value). Each point on the
                        plot presents the median of the bias of all stars having this
                        $T_{\rm eff}$ value (the template effective temperatures
                    take on only discrete values).
                        The coloured bar around it gives the uncertainty on the median;
                        in most cases, this uncertainty is smaller than the symbol.
                        The results for cooler stars are also included in this figure
                        to show the difference in quality of the radial velocity determination.
                        The dashed blue lines indicate $\pm 3$~\kms.
                        The numbers at the top show how many stars 
                        correspond to this temperature.
                        The \emph{right panel} shows the precision. The blue symbols
                        give the external uncertainty (dispersion of RVS value
                        minus reference value), while the red symbols give the 
                        internal precision, i.e. the intrinsic uncertainty as derived
                        by the code. The uncertainties on these values are given by the
                        coloured bars; for the higher $T_{\rm eff}$ values, they have been offset slightly
                        in temperature to avoid overlap. The plots include only
                        stars with $\mathtt{grvs\_mag} \le 12$~mag.
                }
                \label{figure dry-run}
        \end{center}
\end{figure*}

  \begin{figure}[h]
        \begin{center}
                \resizebox{\hsize}{!}{\includegraphics[bb=45 52 310 330]{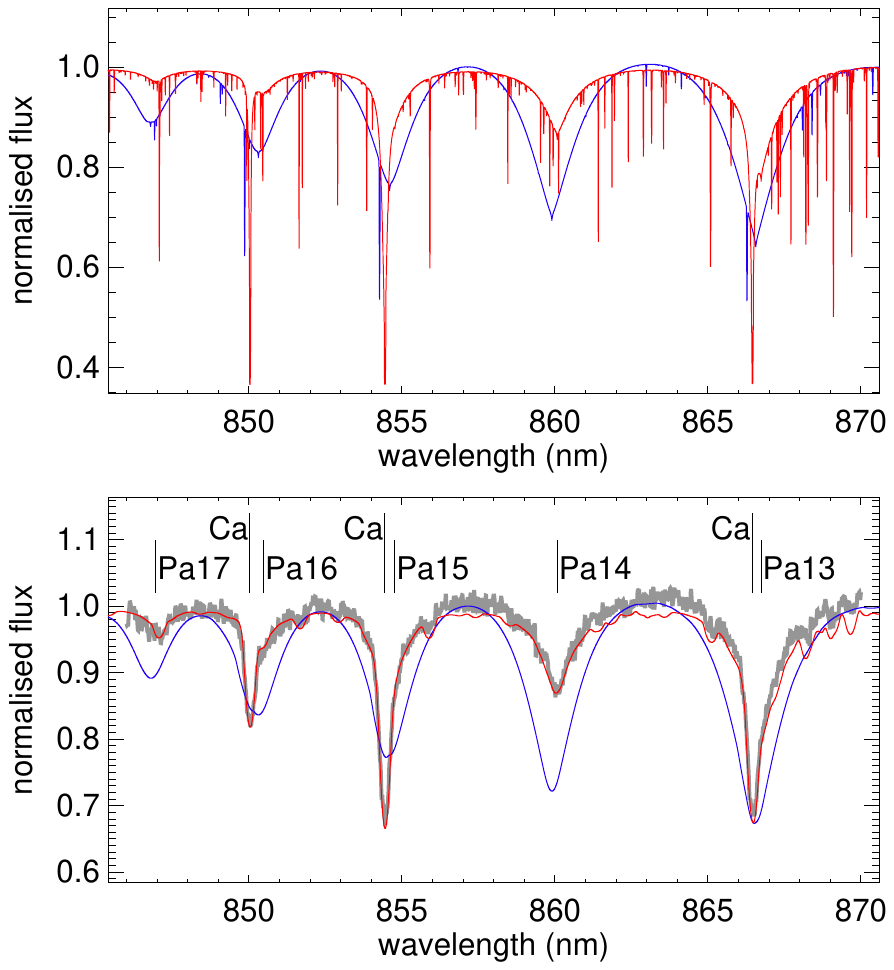}}
                \caption{Example of applying ReDetermineApHotStars.
                        The \emph{top panel} shows the synthetic spectra: the blue curve
                        ($T_{\rm eff} = 11\,500$~K, $\log g = 4.0$~dex, $\mathrm{[M/H]} = -0.50$~dex)
                        presents data to which ReDetermineApHotStars 
                        was not applied,
                        and the red curve shows the best synthetic spectrum as determined by
                        ReDetermineApHotStars
                        ($T_{\rm eff} = 7750$~K, $\log g = 4.5$~dex, $\mathrm{[M/H]} = +0.25$~dex).
                        The \emph{bottom panel} shows the templates derived from these
                        synthetic spectra:
                        they have been convolved
                        with a broadening velocity of 100~\kms (blue curve), 
                        and $80$~\kms (red curve),
                        and with the instrumental profile. They are compared with the
                        observed spectrum of \object{Gaia DR3 51853419339515136}
                        (grey curve).  
                        All spectra are at their vacuum rest wavelength, except for the blue
                        curves, which are shifted by $-60.9$~\kms; this is the offset
                        we found
                        between the case when ReDetermineApHotStars was applied or was left unused.
                        The template selected by ReDetermineApHotStars (red curve) clearly
                        provides a better fit to the observed spectrum and therefore
                        also gives a more correct radial velocity.
                }
                \label{figure example}
        \end{center}
\end{figure}

\section{Method}
\label{section method}

This section describes the method we used to derive the radial
velocities and presents the results of a preliminary run
(Sect.~\ref{section preliminary}). As the results of this preliminary
run were not satisfactory for the hot stars, an improved method was
developed. This is described in Sect.~\ref{section improved method}.

\subsection{STA and MTA}

The processing of the RVS data consists of six pipelines \citep{CU6documentation}; the four main pipelines take care of the spectrum extraction, 
the determination of the
wavelength calibration and the background correction, and the radial
velocity determination. The pipeline we are concerned with consists
of two parts: single-transit analysis (STA) and multi-transit analysis (MTA). These are
described in detail for DR2 in \cite{Sartoretti+18}, and 
the description is updated for DR3 in
\citet{CU6documentation}. Here we provide a summary of the points that are  relevant for this paper.
Figure~\ref{figure STAMTA} also provides an overview.

\subsubsection{STA input}\label{sta input}
The STA part of the pipeline uses the observed spectra as input
that have been wavelength-calibrated and normalised,
and it uses the astrophysical parameters of the most appropriate synthetic 
spectrum\footnote{We use the term `synthetic spectrum' for the 
        spectrum with very high resolution and very high sampling
        that is output from a theoretical
        spectrum synthesis code, and the term `template' when such a
        synthetic spectrum has been convolved with the instrumental profile
        and possibly a line-broadening function (see Sect.~\ref{section STA processing}).}. 
These parameters are the effective temperature ($T_{\rm eff}$),
the gravity ($\log g$), and the metallicity ([M/H]).

In principle, the astrophysical parameters are determined by
Coordination Unit 8 (CU8). However, for DR3, 
the CU8 processing was done after the CU6 processing. The considerable
time needed for the processing and the validation of the results
means that it is not feasible to iterate between the CU6 and CU8 processing. 
CU6 therefore used parameters from the preliminary CU8 run.
This CU8 run used an earlier version of the GSP-Phot \citep{Andrae+22} and 
GSP-Spec \citep{Recio-Blanco+22} pipelines.
GSP-Phot used DR2 photometric data instead of the better-quality and more detailed DR3 BP/RP
data, while GSP-Spec used
RVS spectra from the preliminary CU6 run (discussed in 
Sect.~\ref{section preliminary}).
These initial sets of parameters are thus intrinsically not optimal, and CU6 supplemented these data by other (non-\textit{Gaia}) sources of 
parameters \citep[][their Table 6.4]{CU6documentation}.

The input astrophysical parameters for the choice of the synthetic 
spectrum 
are determined
as follows, in order of priority:
\begin{enumerate}
        \item Parameters from the literature 
        \citep[][their Table 6.4; 3.4 \% of all hot stars]{CU6documentation}.
        \item Parameters from the preliminary run of the CU8
        GSP-Spec pipeline \citep[][3.7 \% of all hot stars]{Recio-Blanco+22}.
        \item Parameters from the preliminary run of the CU8
        GSP-Phot pipeline \citep[][89.6\% of all hot stars]{Andrae+22}.
        \item Parameters determined by CU6 from fitting the RVS spectrum
        with a limited set of stellar templates.
        \citep[DetermineAp --][their Sect.~6.5; 3.3\% of all hot stars]{Sartoretti+18}.
        \item Parameters of the Sun (with corresponding template parameters
        $T_\mathrm{eff} = 5500$, $\log g = 4.5$, [M/H] = 0.0; not applicable to hot stars); 
        as we have no other information
        available for these stars, we use the solar parameters as a default value.
\end{enumerate}
The percentages listed above are applicable for all stars that turn out to have a 
\texttt{rv\_template\_teff}\footnote{We use this 
        font style to indicate 
        the name of columns that are available in the DR3 archive. A
        standard font style is used when we refer to the parameter in general,
        or when we refer to a parameter used in the processing (usually in STA)
        that was not stored
        in the DR3 archive. } $\ge 7000$~K
after the improved method detailed in Sect.~\ref{section improved method} was applied.

Based on the astrophysical parameters, the nearest synthetic spectrum is selected.
The list of available synthetic spectra is given in \cite{RHB-005}. The astrophysical parameters of the synthetic spectrum (which we call template parameters from here on)
are therefore usually slightly different from
the true astrophysical parameters, and in the remainder of this paper, we 
are mainly concerned with the template parameters.

Figure~\ref{figure stats} shows the fraction of the hot stars 
(\texttt{rv\_template\_teff} $\ge 7000$~K)
that obtained
their template parameters from each of the above possibilities.
GSP-Phot is the dominant contributor. 
 GSP-Spec,
which uses the RVS spectra, does not cover the
higher temperatures. This is highly relevant for this paper. The figure also shows information for hot stars
with \texttt{grvs\_mag}
 fainter than magnitude 12, but for reasons explained in 
Sect.~\ref{section improved method}, a cut-off magnitude 12 is
applied for the DR3 archive.

\subsubsection{STA processing}
\label{section STA processing}
For a given transit, the selected synthetic spectrum is converted into a template
by convolving it with the instrumental profile, and possibly with a broadening profile. For the latter, a rotational broadening kernel is used
\citep{2005oasp.book.....G}, with the caveat that the line-broadening velocity can
include other broadening mechanisms as well.
The
template is then normalised, so that it can be compared to the normalised observed spectrum.

The radial velocity and its corresponding uncertainty are
then determined using three different techniques: cross-correlation by
Fourier transform, Pearson correlation, and a $\chi^2$ minimum 
distance \citep[][]{David+14}. 
The spectra from the three CCDs that cover this single transit are processed separately
(because the instrumental profile is different for the three CCDs) to derive a cross-correlation
function (CCF) or a $\chi^2$ function for each spectrum. These three functions are then combined,
and the radial velocity is derived from the maximum of the combined CCF and the minimum
of the combined $\chi^2$ function. For the brighter stars, the broadening 
velocity is
also determined; for details, we refer to \cite{Fremat+22}.

The final value for the transit radial velocity is the median
of the three determinations. During this processing, the spectra are
also checked for the possible
presence of two components and, if they are detected, the corresponding radial velocities
are determined following a dedicated channel
described in \citet{Damerdji+22}.

\subsubsection{MTA processing}
\label{section MTA processing}

In the MTA part of the pipeline, the information of all transits for a given star (with a supposed single-line spectrum) is combined.
This combination is done in a different way for bright stars than for faint stars.
The switch between bright and faint occurs at \texttt{grvs\_mag} $ = 12$~mag, where \texttt{grvs\_mag}
is the magnitude over the RVS bandpass \citep{Sartoretti-GRVS}. 

For the bright stars (\texttt{grvs\_mag} $\le 12$~mag), MTA corrects the STA radial velocities for the barycentric velocity
and takes the median; this quantity is stored in the DR3
archive as 
\texttt{radial\_velocity}. 
The corresponding uncertainty
\texttt{radial\_velocity\_error} is derived 
from the uncertainty on the median (with a floor value of 0.113 \kms); for
details, see \citet[][Sect.~6.4.9]{CU6documentation}.

For the faint stars (\texttt{grvs\_mag} $ = 12 - 14$~mag), MTA uses the STA cross-correlation functions determined by Fourier transform,
shifts them for the barycentric velocity correction, and averages them to obtain
the source cross-correlation function. From the maximum of this function, the 
\texttt{radial\_velocity} is determined. The \texttt{radial\_velocity\_error} is calculated
using the formula of \citet[][Sect.~2.3]{Zucker03}.

In both cases, the MTA processing also provides the spectrum combined over all transits. This 
combined spectrum is shifted by MTA to its rest wavelength.

\subsection{Preliminary radial velocities}
\label{section preliminary}

A preliminary run of the RVS processing pipeline revealed that the 
radial velocities of hot stars show a systematic offset with respect to the reference stars.
The reference stars we use in this comparison are those from the literature listed in \cite{Katz+19}. We also use values
that were derived for cluster members using \textit{Gaia}~DR2.
Although \textit{Gaia}~DR2 contains radial velocities only
for cooler stars ($T_\mathrm{eff} \le 6900$~K), these were used 
by \cite{Soubiran+18} to determine
the cluster radial velocity, and the hotter stars that are members of 
the cluster are assigned the same
velocity as the cluster.

In Fig.~\ref{figure dry-run} we plot the difference between the 
preliminary RVS radial velocities
and the reference values as a function of the template
 $T_{\rm eff}$. As there are a large
number of stars, we condense the information by grouping the stars according to their
template $T_{\rm eff}$ (which takes on only discrete values). For each template
$T_{\rm eff}$, we calculate the median and the uncertainty on the median
of the radial velocity difference, giving us a point and an uncertainty
on the plot. 
The number of stars corresponding to each $T_{\rm eff}$ is listed at the top of the left panel in
Fig.~\ref{figure dry-run}.

Figure~\ref{figure dry-run} shows a clear offset of about $-20$ \kms
for $T_{\rm eff} > 7000$~K. The reason for this offset is a
mismatch between the template and the observed spectrum.
For the $T_{\rm eff}$ range we consider here, the spectra have a mixture
of the calcium infrared triplet and hydrogen Paschen lines. Fig.~\ref{figure hot stars} shows that three of the Paschen lines have a calcium
line on their blueward side. If the template does not have the correct relative
intensities of the calcium and Paschen lines, the radial velocity determination
techniques will attempt to shift the theoretical Paschen lines 
towards the position of the
observed calcium lines, leading to an incorrect, blue-shifted, velocity.
A template without template mismatch would not have this problem, and the
correct radial velocity would be obtained.

The right panel of Fig.~\ref{figure dry-run} shows two versions of the precision.
The blue symbols give the external precision:
this is the half-range between the 15.85 and 84.15 percentile of the distribution of the radial velocity differences
\citep[see][their Eqs. 4-6]{Katz+19}. 
The red symbols give the internal precision: this is the median value
(and its uncertainty) of the radial velocity
uncertainty (\texttt{radial\_velocity\_error})
as derived in the MTA processing (Sec.~\ref{section MTA processing}). The precision of stars hotter than 
6500~K is considerably worse than the precision for stars at lower temperatures. The internal
precision is lower than the external one, indicating that the 
radial velocity uncertainty is underestimated for these stars.
We also checked the influence of the uncertainty of the reference values
on the external precision, but this is negligible compared
to the difference between external and internal precision.

\begin{figure*}
        \begin{center}
                \includegraphics[bb=25 20 530 330]{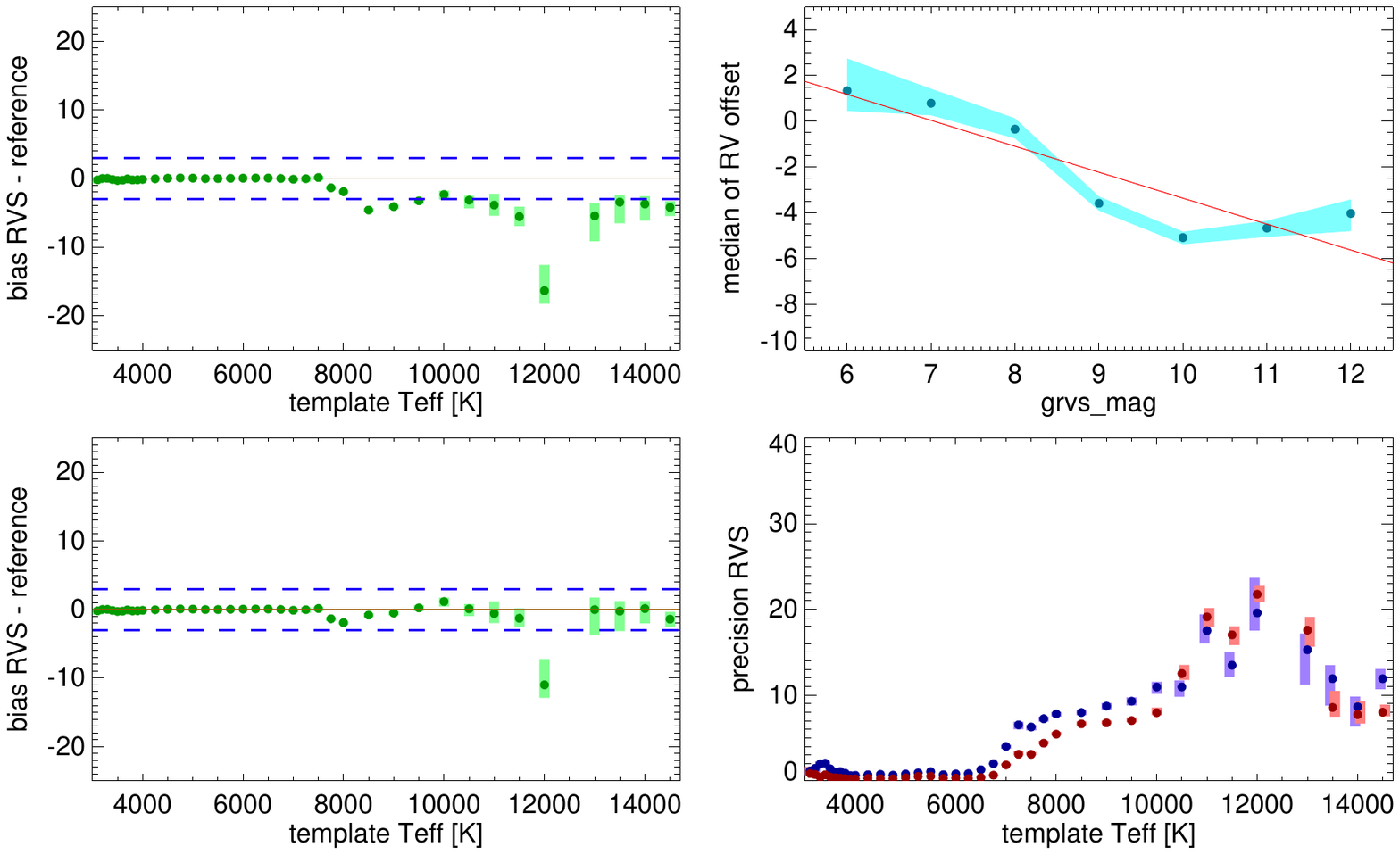}
                \caption{Radial velocities compared to their reference 
                        values, where ReDetermineApHotStars has been applied to 
                        the hotter stars.
                        The \emph{top left panel} shows the bias of the radial velocity of
                        the stars (RVS value minus reference value). It should be
                        compared to the equivalent plot in Fig.~\ref{figure dry-run} to
                        see the improvement due to ReDetermineApHotStars.
                        The dashed blue lines indicate $\pm 3\kms$.
                        The \emph{top right panel} plots the median of the 
                        radial velocity bias of stars in the $T_{\rm eff}=8500-14\,500$~K range,
                        binned in one-magnitude-wide ({\tt grvs\_mag}) bins. 
                        The relation between the two is fitted linearly (red line).
                        Applying the magnitude-related correction leads to an improved
                        set of radial velocities, which are once again compared
                        to their reference values (\emph{bottom left panel}).
                        The precision of these improved values is shown in 
                        the \emph{bottom right panel}. 
                        Similarly to Fig.~\ref{figure dry-run}, the blue symbols
                        give the external precision, and the red symbols 
                        give the internal precision. 
                        All plots include only
                        stars with $\mathtt{grvs\_mag} \le 12$~mag.
                }
                \label{figure magnitude effect}
        \end{center}
\end{figure*}

  \begin{figure}[h]
        \begin{center}
                \resizebox{\hsize}{!}{\includegraphics[]{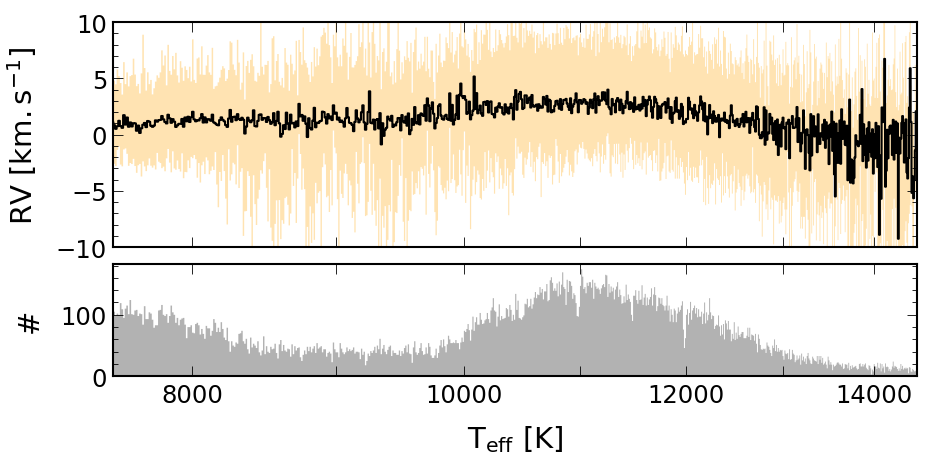}}
                \caption{
Radial velocity shift that had to be applied by ESP-HS in
CU8 to align the theoretical spectrum to the observed RVS spectrum
as determined by MTA. 
It is plotted as a function of the $T_{\rm eff}$ determined by ESP-HS.
The top panel shows the running median (computed over 
consecutive $T_{\rm eff}$ bins of 10~K) and the interquantile
dispersion (15.85 \% -- 84.15 \%) of this offset as a function of the effective temperature. 
The bottom panel shows how many stars went into each 10~K bin.          
        }
                \label{figure further test}
        \end{center}
\end{figure}

\section{Improved method}
\label{section improved method}

Because of the unsatisfactory results of the preliminary run for hot stars,
we decided on a new approach, implemented as the module ReDetermineApHotStars. This module does an exhaustive brute-force search among the combination of many
templates and broadening velocities to find the combination 
that best fits (in $\chi^2$ sense) 
all transit spectra of a given star.
As part of the procedure, ReDetermineApHotStars also shifts the template being explored in velocity
space, and it thus also determines the transit radial velocities. However, only
the updated template parameters and the broadening velocity are output from ReDetermineApHotStars; 
the radial velocity determination is left to the three
radial velocity modules (Sect.~\ref{section STA processing}).

The way the ReDetermineApHotStars module is included in the STA part of the
pipeline is shown in Fig.~\ref{figure STAMTA}.
ReDetermineApHotStars is called for stars with a template $T_{\rm eff} \ge 7000$~K. It is also called for $6500 \le T_{\rm eff} < 7000$~K when the atmospheric parameters are not from literature data (see Sect.~\ref{sta input}). We apply this procedure
only to stars that are bright enough ($G_{\rm RVS} \le 12$~mag) so that we 
have a sufficient S/N in the spectra.

The $G_{\rm RVS}$ value used in this procedure is not exactly the same as the
\texttt{grvs\_mag}
available in the DR3 archive. The latter value is determined in the
MTA part of the pipeline as the
median of all valid transit $G_{\rm RVS}$ values \citep{Sartoretti-GRVS}.
During the run of STA, this median value was not yet available, therefore STA
used the first valid value from the list of transits.

Because of the requirement for bright enough stars for ReDetermineApHotStars,
the hot-star radial velocities in the DR3 archive are mostly limited to
\texttt{grvs\_mag} $\le 12$~mag. The few stars that are fainter than this 
are due to the difference between the value used by STA and MTA, as 
explained above.

\begin{table}
        \caption{Parameters of the grid of synthetic spectra used in ReDetermineApHotStars.}
        \label{table synthetic spectra}
\centering
\begin{tabular}{lccc}
        \hline
        \hline
        Code & Parameter &  Range &  Step \\
        \hline
        MARCS & $T_{\rm eff}$ [K] & 6500 .. 8000 & 250 \\
              & log $g$ [dex] & $-0.5$ ..  $+5.0$ & 0.5  \\
              & [M/H] [dex] & $ -5.0$ .. $-3.0$ & 1.0 \\
              &       & $ -2.5$ .. $-1.0 $ & 0.5 \\
              &       & $ -0.75$ .. $+1.0 $ & 0.25 \\
        A stars& $T_{\rm eff}$ [K] & 8500 .. 12\,000 & 500 \\
              & $T_{\rm eff}$ [K] & 13\,000 .. 14\,500 & 500 \\
              & log $g$ [dex] & $+2.0$ .. $+5.0 $ & 0.5  \\
              & [M/H] [dex] & $-0.5$ .. $+0.25$ & 0.25 \\
        \hline
\end{tabular}

\end{table}

Because the brute-force grid search in ReDetermineApHotStars is computing-time intensive,
various optimisations are introduced in that module. 
We combine the observed spectra from the
three CCDs into a single spectrum, so that we have to handle fewer spectra.
For the combination, we use a wavelength grid
with a constant step in $\log \lambda$, so that the radial velocity shifts can be
made by simple index manipulation. The 
templates we compare with have been pre-calculated. We base them on 
the synthetic spectra used in the RVS processing \citep{RHB-005,CU6documentation},
selecting all those
that have a $T_{\rm eff} \ge 6500$~K. 
The parameter ranges of the synthetic spectra we use are listed in
Table~\ref{table synthetic spectra}. We convolve each synthetic spectrum
with a pre-specified grid of rotational velocities and with the instrumental
profile consisting of a Gaussian with the nominal RVS resolving
power of 11\,500. The templates are also on a wavelength grid
with a constant step in log $\lambda$.

ReDetermineApHotStars has a main loop over the set of the pre-calculated templates and an 
inner loop over all transits of a given star. In the inner loop, it 
determines the best radial velocity for each transit. It does
so by exploring a range of radial velocities and calculating the $\chi^2$
between observed spectrum and template. As part of this procedure, the normalisation of the observed spectrum is also slightly adapted to agree
as well as possible with the normalisation of the template. 
Again, for optimisation purposes, we first use a coarse velocity grid to obtain 
an approximate radial velocity from the minimum $\chi^2$. The velocity grid is then
refined around this approximate result to obtain a better value, which is then
even further refined by parabolic interpolation. In this way, we obtain the best radial velocity for a specific combination of pre-calculated template and transit.
The inner loop then loops over all transits, and 
the best-fitting pre-calculated template for the star is then determined by the minimum of the $\chi^2$ summed over all transits of that star.

As a further optimisation, we ordered the pre-calculated templates in such a way that we first explore
the templates that show the highest difference among themselves (the difference between the templates is measured by the sum of the square of the flux differences).
In this way, we quickly determine a template 
with a reasonably good $\chi^2$ value for the star.
When we proceed with the set of pre-calculated templates, we 
find many templates with a much higher $\chi^2$ than the best $\chi^2$ obtained
thus far. It therefore frequently happens that as we loop over the transits, we already arrive at a summed $\chi^2$ 
that is too large before we have processed all the transits of that star. 
We can therefore
break off further processing of this specific template in the inner loop, 
and in this way reduce the required computing time. 

An example of applying ReDetermineApHotStars is shown in
Fig.~\ref{figure example}. 
ReDetermineApHotStars works on the separate transit spectra, but for the example,
we chose an RVS spectrum from the \textit{Gaia} DR3 archive, which is a combination
of all its transit spectra. As for all archive spectra, this spectrum was
shifted to its rest wavelength, in this case, using the radial velocity 
based on the processing by ReDetermineApHotStars.
The observed spectrum (shown in grey in the
bottom panel) contains a mixture of Paschen and calcium lines. 
This spectrum was then compared to two templates that were derived from the
synthetic spectra shown in the top panel. The
template used in the preliminary run has a relatively high effective temperature 
($T_{\rm eff} = 11\,500$~K) and 
therefore shows only weak calcium lines. When fitting such a template
(blue curve)
to the observed spectrum, the theoretical Paschen lines are blue-shifted
in attempting to partially fit the observed calcium lines. 
This gives
an incorrect, blue-shifted, radial velocity of $-60.9$~\kms.
The bottom panel shows that the blue curve fits the region around Pa13 and Pa15
reasonably well, but clearly fails around Pa14, Pa16, and Pa17.
ReDetermineApHotStars finds 
a much cooler template that has the appropriate mixture of Paschen
and calcium lines, and therefore fits the observed spectrum much better.
Using this template results in a more correct radial velocity. This
radial velocity was used to place the DR3 archive spectrum 
(shown in Fig.~\ref{figure example}) at its
rest wavelength.

\section{Results}
\label{section results}

\subsection{Improved radial velocities}

The top left panel of Fig.~\ref{figure magnitude effect} shows the improvement in
the bias of the radial velocities (RVS radial velocity minus the reference
value). The offset of about $-20\kms$ shown in Fig.~\ref{figure dry-run}
is reduced to about $-5\kms$ for most of the effective temperature range.

The remaining $-5\kms$ offset can be further reduced by considering
the plot in the top right panel
of Fig.~\ref{figure magnitude effect}. When we plot the bias of the radial
velocities (for $T_{\rm eff}=8500-14\,500$~K stars)
as a function of magnitude ({\tt grvs\_mag}), a clear dependence 
on magnitude is seen. We approximate this dependence with a linear fit,
\begin{equation}
\mathrm{RV_{offset}} = 7.98 - 1.135 * \mathtt{grvs\_mag}.
\label{equation offset}
\end{equation}

After we apply Eq.~\ref{equation offset} to correct the radial velocities, we
can again plot their bias (Fig.~\ref{figure magnitude effect}, bottom-left panel).
Now, almost all points show a bias lower than 3\kms. The only exception is $T_{\rm eff}=12\,000$~K; we were unable to determine why this temperature
behaves somewhat differently. It is important to realise that 
the \texttt{radial\_velocity} listed in the DR3 archive
does not include this correction.
It is up to the user of the DR3 hot-star radial velocities 
to decide whether they
wish to apply this correction. As an example, we refer to
\cite{Drimmel+22}, where this correction is applied.

The magnitude-dependent effect (Fig.~\ref{figure magnitude effect}, top right panel)
is due to the difference in normalisation between
the template and the observed spectrum. The normalisation
consists of repeatedly fitting a second-degree
polynomial to the fluxes and rejecting flux values that deviate
too much from the polynomial. A stricter cutoff is used for fluxes below the
polynomial than for those above it. Although the same technique is used
for both the template and the observation, the presence of noise in
the observation leads to a polynomial with a somewhat different slope than
that of the noise-free template. 
With increasing effective temperature, the wider Paschen lines start to
dominate, and radial velocity measurements become more sensitive to this slope mismatch, leading to a systematic offset.
The effect becomes larger for the fainter stars because they have noisier observed spectra. For DR4 we are exploring a different
normalisation technique where the normalised template is used as a reference
for the normalisation of the observed spectrum. Preliminary indications are
that this substantially reduces the magnitude-dependent effect described here.

The bottom right panel of Fig.~\ref{figure magnitude effect} shows the
two versions of the precision in the same way as for 
Fig.~\ref{figure dry-run}.
With the improved method, both curves are now lower, indicating 
the improved precision. Furthermore, the internal
precision (red curve) is now much closer to the external one.
As for Fig.~\ref{figure dry-run}, we examined the influence of the uncertainty of the reference values
on the external precision and found that it is negligible compared
to the difference between external and internal precision.
From the comparison of the two curves, we deduce that the
\texttt{radial\_velocity\_error} is underestimated in the
$T_{\rm eff}=8000 - 10\,000$~K range by $\sim$\,30\%.
For higher temperatures ($T_{\rm eff}=10\,500 - 13\,000$~K), 
the \texttt{radial\_velocity\_error} is slightly overestimated
(though the uncertainty regions of the two precisions partially overlap).

Figure~\ref{figure magnitude effect} does not plot results for stars hotter than 14\,500~K. These stars still present a considerable offset that could
not be solved in DR3 and will be handled in DR4.

\begin{figure}
        \begin{center}
                \resizebox{\hsize}{!}{\includegraphics[bb=45 52 310 300]{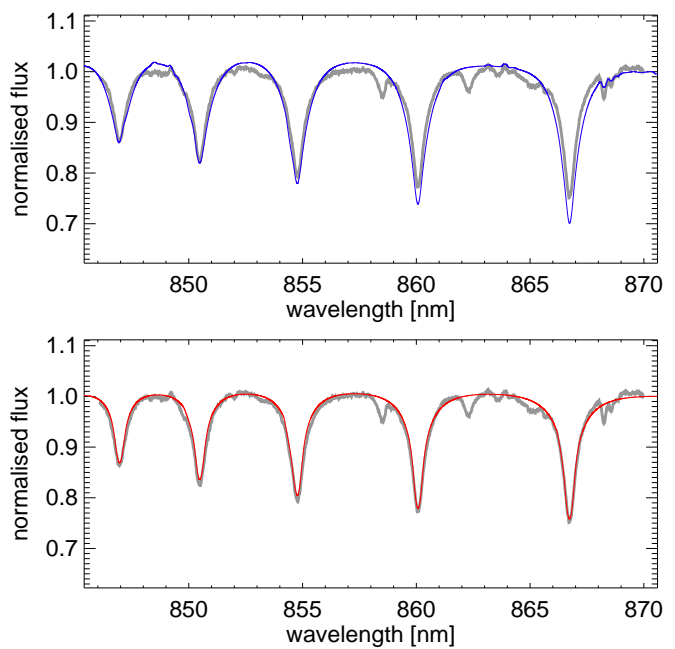}}
                \caption{Degeneracy between $T_{\rm eff}$ and metallicity for HDE~250290.
                        The \emph{top panel} shows a fit of the observed spectrum (grey)
                        with a template (blue curve) corresponding 
                        closely to the
                        parameters of \citet[][$T_{\rm eff} = 16\,000$~K, $\log g = 2.5$~dex, $\mathrm{[M/H]} = 0.0$~dex, and line-broadening velocity
                        $v_\mathrm{broad}=50$~\kms]{SimonDiaz+17}.
                        The \emph{bottom panel} shows the fit with the template (red curve) as
                        determined by ReDetermineApHotStars
                        ($T_{\rm eff} = 6750$~K, $\log g = 2.5$~dex, $\mathrm{[M/H]} = -5.0$~dex,
 and                        $v_\mathrm{broad}=80$~\kms).
                }
                \label{figure degeneracy}
        \end{center}
\end{figure}

\begin{figure}
        \begin{center}
                \resizebox{\hsize}{!}{\includegraphics[bb=5 5 290 220]{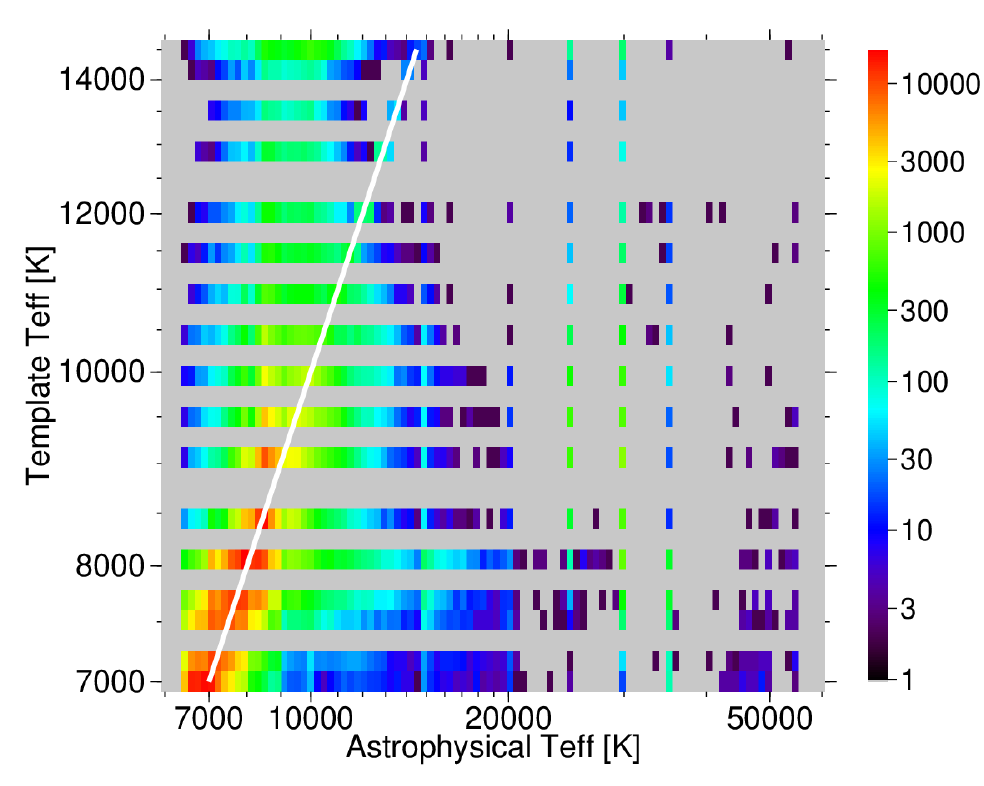}}
                \caption{Template $T_{\rm eff}$ as a function of the input
                        astrophysical $T_{\rm eff}$. The colour scale gives the
                        numbers of stars (on a logarithmic scale). The data
                        are limited to those with $\mathtt{grvs\_mag} \le 12$~mag.
                        The diagonal is indicated with a solid white line.
                }
                \label{figure params}
        \end{center}
\end{figure}

\begin{figure*}
        \begin{center}
                \includegraphics[bb=30 20 530 480]{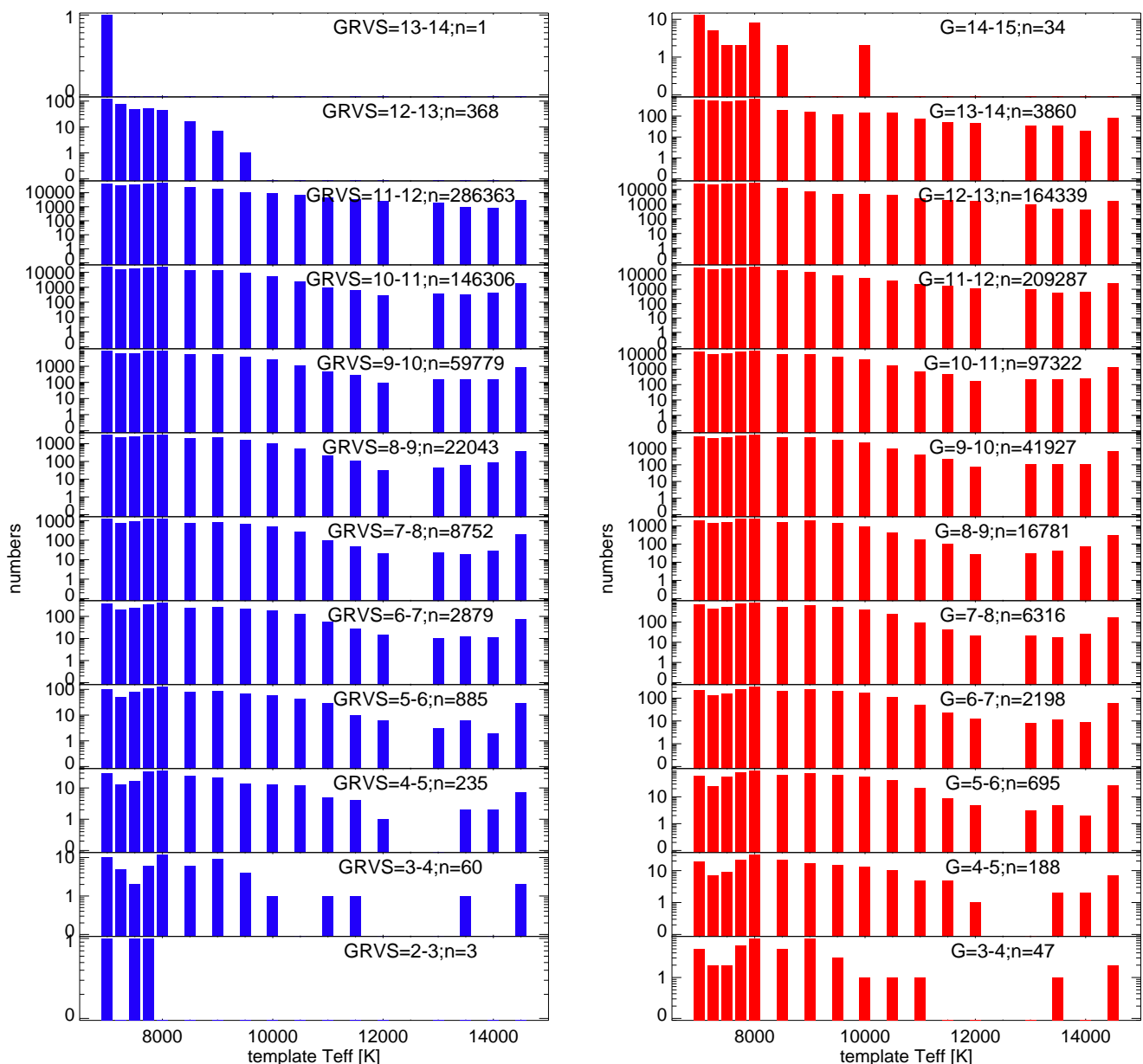}
                \caption{Number of hot stars (log scale) with radial velocities 
                        in DR3, split up according to
                        \texttt{rv\_template\_teff},
                        \texttt{grvs\_mag} (left panel) and 
                        $G$ magnitude (\texttt{phot\_g\_mean\_mag})  (right panel).
                        Each sub-panel lists the magnitude range and the number of stars
                        within that range.
                        While DR3 has a formal cutoff of magnitude 12, there are some stars
                        fainter than this because of the difference between $G_\mathrm{RVS}$
                        and \texttt{grvs\_mag} (see Sect.~\ref{section improved method}).
                }
                \label{figure numbers}
        \end{center}
\end{figure*}

A further test of the hot-star radial velocities is provided by
the CU8 processing to determine the astrophysical parameters
from the spectra delivered by CU6. As part of the astrophysical parameter
determination performed by the Extended Stellar Parameterizer - Hot Stars 
\citep[ESP-HS;][]{DR3-DPACP-157}, the spectra are compared to synthetic 
spectra. When RVS data are available, the ESP-HS performs a first estimation 
of the astrophysical parameters of hot stars from the analysis of BP/RP data and then applies
a cross-correlation technique in Fourier space to measure any remaining radial
velocity offset of the RVS spectrum relative to the corresponding interpolated 
theoretical one. The RVS spectra used by CU8 were shifted to their rest wavelength by the
CU6 processing (Sect.~\ref{section MTA processing}), 
therefore no offset should remain in principle. This CU6 shift
is based on the combination of the 
transit radial velocities, however, while the ESP-HS test looks at the single combined
spectrum for the source. This can result in a somewhat different radial velocity, and thus 
the ESP-HS test provides an independent way of validating the correctness of the CU6 radial
velocity determination.

The ESP-HS module applies the radial velocity offset it measures to proceed 
further with the astrophysical parameter determination by fitting both BP/RP and RVS data. 
While the value of this CU8 offset shift is not published in DR3, it is 
kept for validation purposes and provides an a posteriori check on the 
CU6 results. An interesting point of this test is that it can be applied to 
a much larger number of stars, as we do not need a reference value for the radial
velocity. Figure~\ref{figure further test} shows the variation in required shift with the effective
temperature measured by ESP-HS. Its median computed between 7500~K and 14\,500~K is $+$1.7~\kms
, with a half interquantile dispersion of 5.9~\kms. These values are low and are
consistent with the results found from the comparison with the reference values 
(Fig.~\ref{figure magnitude effect}).
The CU8 results thus confirm that the combined spectra are in the rest frame and that the CU6 determined radial velocities are accurate to within the claimed $\pm 3$~\kms.
The median here has the opposite sign from that
in Fig.~\ref{figure magnitude effect} because this test measures the
required shift of theoretical spectrum to align it to the observed spectrum.

\subsection{Astrophysical versus template parameters}

The improved selection of the best template does
not necessarily mean that the DR3 template parameters
(\texttt{rv\_template\_teff}, 
\texttt{rv\_template\_logg}, 
and \texttt{rv\_template\_fe\_h} in the archive) 
describe the spectral type well. One effect that is responsible for this
is the degeneracy between $T_{\rm eff}$ and metallicity in the very short
wavelength range covered by RVS.

As an example of this degeneracy, Fig.~\ref{figure degeneracy}
shows the RVS archive spectrum of the B3 Ib star Gaia DR3 3424656293035086208 = \object{HDE 250290}. The top panel fits the observation with a template that is as
close as possible to the parameters listed by \citet{SimonDiaz+17}. The bottom
plot uses the template from ReDetermineApHotStars. The 
\cite{SimonDiaz+17} values are $T_{\rm eff} = 16\,000$~K, and we assume
they used $\mathrm{[M/H]} = 0.0$, as is appropriate for this hot Galactic star.
The ReDetermineApHotStars result has a much lower $T_{\rm eff}$ ($6750$~K)
and an extremely low metallicity ($\mathrm{[M/H]} = -5.0$). Based on the
RVS data alone, as shown in Fig.~\ref{figure degeneracy}, the cooler template with the very
low metallicity fits better. Only by using non-RVS
data (as done by \cite{SimonDiaz+17}) does it become clear that the
$\mathrm{[M/H]} = -5.0$ metallicity 
does not describe the atmospheric
parameters of this star correctly.

\begin{figure}
        \begin{center}
                \resizebox{\hsize}{!}{\includegraphics[bb=32 150 755 530]{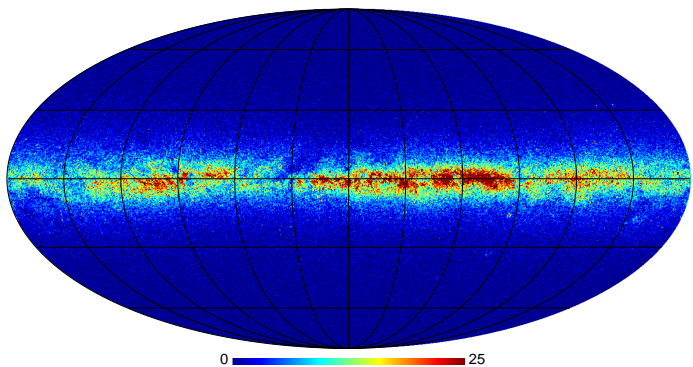}}
                \caption{Distribution on the sky (in Galactic coordinates) of the 
                        approximately half a million hot stars  
                        with DR3 radial velocities. The colour scale gives the number
                        of stars per HEALpix of level 7.
                }
                \label{figure healpix}
        \end{center}
\end{figure}

A further illustration of the difference between 
the template parameters as used in the radial velocity 
determination and the 
astrophysical parameters
is shown in Fig.~\ref{figure params}. The astrophysical parameters
used for this figure are the input parameters discussed 
in Sect.~\ref{sta input}.
The colour scale shows the number of stars having a specific combination of
astrophysical and template $T_{\rm eff}$. Many stars lie close
to the diagonal, and the two temperatures agree well. Some stars have a high astrophysical $T_{\rm eff}$, however, where ReDetermineApHotStars
has assigned a low template $T_{\rm eff}$. Conversely, for stars
with an astrophysical $T_{\rm eff} \approx 10\,000$~K,  ReDetermineApHotStars
sometimes finds a higher $T_{\rm eff}$. We stress again that the
sole purpose of the template $T_{\rm eff}$ is to obtain the best possible
radial velocity; it does not necessarily describe the astrophysical
properties of the star well.

\subsection{Hot stars with cool templates}

As mentioned in Sect.~\ref{sta input},
we were only able to use a preliminary version of the CU8 astrophysical parameters as input
data. However, towards the end of the validation phase of the CU6 results, the 
CU8 results were available to us. These include the astrophysical parameters
determined by the
ESP-HS \citep[][]{DR3-DPACP-157}. 
In these data,
we found a number of stars for which
we used a cool-star template for a star that is now
classified by CU8 as a hot star.

This can again lead to a systematic offset because the calcium lines
in the cool-star template try to fit the Paschen lines in the hot-star observed
spectrum. We therefore carefully examined these cases again to determine which provided incorrect radial velocities.
On the basis of this, the radial velocities of 20\,470 hot stars 
(i.e. 3.8\,\% of all hot stars) with 
cool templates were removed
from the DR3 archive.

\subsection{Number of hot stars}

The important difference between DR2 and DR3 discussed in this paper
is the extension to 14 500 K of the effective temperature range for which 
radial velocities are determined.
Of the total number of 33\,812\,183 of stars with radial velocities in DR3, 543\,017 lie in the range 
\texttt{rv\_template\_teff} = 7000 -- 14\,500~K. 
The distribution of these hot stars with 
\texttt{rv\_template\_teff}, \texttt{grvs\_mag} and 
$G$ magnitude (\texttt{phot\_g\_mean\_mag}) is shown in
Fig.~\ref{figure numbers}.

As expected, most of these stars are in the cooler, fainter part of
the diagram. We introduced a formal cutoff in magnitude of 
\texttt{grvs\_mag} $=12$~mag,
so that the S/N of the spectra is high enough for ReDetermineApHotStars
to give good results. Some
stars that are fainter than this cutoff are shown: this is due to the
(sometimes) different value that was used by STA or MTA
(see Sect.~\ref{section improved method}).
Figure~\ref{figure numbers} has no stars with a template of 12\,500~K because
our list does not contain a synthetic spectrum with this $T_{\rm eff}$.

Figure~\ref{figure healpix} shows the distribution on the sky of the
hot stars whose radial velocity is listed in DR3. As expected,
the approximately half a million stars are concentrated in the disk
of our Galaxy.

\section{Conclusions}
\label{section conclusions}

The third \textit{Gaia} data release (\textit{Gaia} DR3) is the first
data release to contain radial velocities of stars with a template $T_{\rm eff}$
(\texttt{rv\_template\_teff})
between 7000 and 14\,500~K. The archive contains 543\,017 stars 
in this range.

A preliminary run of the pipeline
deriving these hot-star radial velocities from the spectra of the Radial Velocity
Spectrometer gave unsatisfactory results. The problem is due to the proximity
to and systematic blueward offset
of the calcium infrared triplet lines to the Paschen hydrogen lines.
This leads 
to a systematic offset in radial velocity if the template does not
describe the observed spectrum correctly. For this reason, a specific 
module was developed that improves the choice of the template. Using
this module leads to a much better agreement between the \textit{Gaia} radial
velocities and the reference data. As a sufficiently high S/N is required
for the module to be applied, the DR3 archive mostly contains only
hot stars with \texttt{grvs\_mag} $\le 12$~mag.

The user of the \textit{Gaia} DR3 hot-star radial velocities should be
aware of the following caveats.
        The template parameters $T_{\rm eff}$, $\log g$, and [M/H] used by CU6 (\texttt{rv\_template\_teff}, \texttt{rv\_template\_logg}, and
        \texttt{rv\_template\_fe\_h}
        in the archive) are not necessarily a good description of the spectral
        type of the star.
        A magnitude-dependent effect is still present in the hot-star radial velocities of the \textit{Gaia}
        DR3 archive. 
        For stars with $8500 \le \mathtt{rv\_template\_teff} \le 14\,500$~K 
        and $6 \le \mathtt{grvs\_mag} \le 12$~mag, this effect can be corrected for 
        as follows:
        \begin{equation}
        \mathrm{RV_{corrected}} = \mathtt{radial\_velocity}
        - 7.98 + 1.135 * \mathtt{grvs\_mag}, \nonumber
        \end{equation}
        where \texttt{rv\_template\_teff}, \texttt{grvs\_mag}, and
                \texttt{radial\_velocity} are the quantities listed in the DR3 archive.
                This correction has been derived based on the median offsets of the
                DR3 radial velocities and the reference data. Individual stars can behave quite differently.

\section*{Acknowledgements\label{sec:acknowl}}
\addcontentsline{toc}{chapter}{Acknowledgements}
This work presents results from the European Space Agency (ESA) space mission \gaia. \gaia\ data are being processed by the \gaia\ Data Processing and Analysis Consortium (DPAC). Funding for the DPAC is provided by national institutions, in particular the institutions participating in the \gaia\ MultiLateral Agreement (MLA). The \gaia\ mission website is \url{https://www.cosmos.esa.int/gaia}. The \gaia\ archive website is \url{https://archives.esac.esa.int/gaia}.
Acknowledgements are given in Appendix~\ref{ssec:appendixA}.
We thank Rosanna Sordo for comments on a preliminary version of this paper.
This work has used the following software products:
IDL (\url{https://www.l3harrisgeospatial.com/Software-Technology/IDL}),
HEALPix (\url{https://healpix.sourceforge.io/}),
Matplotlib \citep[][\url{https://matplotlib.org}]{Hunter:2007},
SciPy \citep[][\url{https://www.scipy.org}]{2020SciPy-NMeth}, and
NumPy \citep[][\url{https://numpy.org}]{harris2020array}.  
This research has made use of the SIMBAD database, operated at CDS, Strasbourg, France.

\bibliographystyle{aa}
\bibliography{43685}

\begin{appendix}
\section{}\label{ssec:appendixA}
This work presents results from the European Space Agency (ESA) space mission \gaia. \gaia\ data are being processed by the \gaia\ Data Processing and Analysis Consortium (DPAC). Funding for the DPAC is provided by national institutions, in particular the institutions participating in the \gaia\ MultiLateral Agreement (MLA). The \gaia\ mission website is \url{https://www.cosmos.esa.int/gaia}. The \gaia\ archive website is \url{https://archives.esac.esa.int/gaia}.

The \gaia\ mission and data processing have financially been supported by, in alphabetical order by country:
\begin{itemize}
\item the Algerian Centre de Recherche en Astronomie, Astrophysique et G\'{e}ophysique of Bouzareah Observatory;
\item the Austrian Fonds zur F\"{o}rderung der wissenschaftlichen Forschung (FWF) Hertha Firnberg Programme through grants T359, P20046, and P23737;
\item the BELgian federal Science Policy Office (BELSPO) through various PROgramme de D\'{e}veloppement d'Exp\'{e}riences scientifiques (PRODEX) grants and the Polish Academy of Sciences - Fonds Wetenschappelijk Onderzoek through grant VS.091.16N, and the Fonds de la Recherche Scientifique (FNRS), and the Research Council of Katholieke Universiteit (KU) Leuven through grant C16/18/005 (Pushing AsteRoseismology to the next level with TESS, GaiA, and the Sloan DIgital Sky SurvEy -- PARADISE);  
\item the Brazil-France exchange programmes Funda\c{c}\~{a}o de Amparo \`{a} Pesquisa do Estado de S\~{a}o Paulo (FAPESP) and Coordena\c{c}\~{a}o de Aperfeicoamento de Pessoal de N\'{\i}vel Superior (CAPES) - Comit\'{e} Fran\c{c}ais d'Evaluation de la Coop\'{e}ration Universitaire et Scientifique avec le Br\'{e}sil (COFECUB);
\item the Chilean Agencia Nacional de Investigaci\'{o}n y Desarrollo (ANID) through Fondo Nacional de Desarrollo Cient\'{\i}fico y Tecnol\'{o}gico (FONDECYT) Regular Project 1210992 (L.~Chemin);
\item the National Natural Science Foundation of China (NSFC) through grants 11573054, 11703065, and 12173069, the China Scholarship Council through grant 201806040200, and the Natural Science Foundation of Shanghai through grant 21ZR1474100;  
\item the Tenure Track Pilot Programme of the Croatian Science Foundation and the \'{E}cole Polytechnique F\'{e}d\'{e}rale de Lausanne and the project TTP-2018-07-1171 `Mining the Variable Sky', with the funds of the Croatian-Swiss Research Programme;
\item the Czech-Republic Ministry of Education, Youth, and Sports through grant LG 15010 and INTER-EXCELLENCE grant LTAUSA18093, and the Czech Space Office through ESA PECS contract 98058;
\item the Danish Ministry of Science;
\item the Estonian Ministry of Education and Research through grant IUT40-1;
\item the European Commission’s Sixth Framework Programme through the European Leadership in Space Astrometry (\href{https://www.cosmos.esa.int/web/gaia/elsa-rtn-programme}{ELSA}) Marie Curie Research Training Network (MRTN-CT-2006-033481), through Marie Curie project PIOF-GA-2009-255267 (Space AsteroSeismology \& RR Lyrae stars, SAS-RRL), and through a Marie Curie Transfer-of-Knowledge (ToK) fellowship (MTKD-CT-2004-014188); the European Commission's Seventh Framework Programme through grant FP7-606740 (FP7-SPACE-2013-1) for the \gaia\ European Network for Improved data User Services (\href{https://gaia.ub.edu/twiki/do/view/GENIUS/}{GENIUS}) and through grant 264895 for the \gaia\ Research for European Astronomy Training (\href{https://www.cosmos.esa.int/web/gaia/great-programme}{GREAT-ITN}) network;
\item the European Cooperation in Science and Technology (COST) through COST Action CA18104 `Revealing the Milky Way with \gaia (MW-Gaia)';
\item the European Research Council (ERC) through grants 320360, 647208, and 834148 and through the European Union’s Horizon 2020 research and innovation and excellent science programmes through Marie Sk{\l}odowska-Curie grant 745617 (Our Galaxy at full HD -- Gal-HD) and 895174 (The build-up and fate of self-gravitating systems in the Universe) as well as grants 687378 (Small Bodies: Near and Far), 682115 (Using the Magellanic Clouds to Understand the Interaction of Galaxies), 695099 (A sub-percent distance scale from binaries and Cepheids -- CepBin), 716155 (Structured ACCREtion Disks -- SACCRED), 951549 (Sub-percent calibration of the extragalactic distance scale in the era of big surveys -- UniverScale), and 101004214 (Innovative Scientific Data Exploration and Exploitation Applications for Space Sciences -- EXPLORE);
\item the European Science Foundation (ESF), in the framework of the \gaia\ Research for European Astronomy Training Research Network Programme (\href{https://www.cosmos.esa.int/web/gaia/great-programme}{GREAT-ESF});
\item the European Space Agency (ESA) in the framework of the \gaia\ project, through the Plan for European Cooperating States (PECS) programme through contracts C98090 and 4000106398/12/NL/KML for Hungary, through contract 4000115263/15/NL/IB for Germany, and through PROgramme de D\'{e}veloppement d'Exp\'{e}riences scientifiques (PRODEX) grant 4000127986 for Slovenia;  
\item the Academy of Finland through grants 299543, 307157, 325805, 328654, 336546, and 345115 and the Magnus Ehrnrooth Foundation;
\item the French Centre National d’\'{E}tudes Spatiales (CNES), the Agence Nationale de la Recherche (ANR) through grant ANR-10-IDEX-0001-02 for the `Investissements d'avenir' programme, through grant ANR-15-CE31-0007 for project `Modelling the Milky Way in the \gaia era’ (MOD4Gaia), through grant ANR-14-CE33-0014-01 for project `The Milky Way disc formation in the \gaia era’ (ARCHEOGAL), through grant ANR-15-CE31-0012-01 for project `Unlocking the potential of Cepheids as primary distance calibrators’ (UnlockCepheids), through grant ANR-19-CE31-0017 for project `Secular evolution of galxies' (SEGAL), and through grant ANR-18-CE31-0006 for project `Galactic Dark Matter' (GaDaMa), the Centre National de la Recherche Scientifique (CNRS) and its SNO \gaia of the Institut des Sciences de l’Univers (INSU), its Programmes Nationaux: Cosmologie et Galaxies (PNCG), Gravitation R\'{e}f\'{e}rences Astronomie M\'{e}trologie (PNGRAM), Plan\'{e}tologie (PNP), Physique et Chimie du Milieu Interstellaire (PCMI), and Physique Stellaire (PNPS), the `Action F\'{e}d\'{e}ratrice \gaia' of the Observatoire de Paris, the R\'{e}gion de Franche-Comt\'{e}, the Institut National Polytechnique (INP) and the Institut National de Physique nucl\'{e}aire et de Physique des Particules (IN2P3) co-funded by CNES;
\item the German Aerospace Agency (Deutsches Zentrum f\"{u}r Luft- und Raumfahrt e.V., DLR) through grants 50QG0501, 50QG0601, 50QG0602, 50QG0701, 50QG0901, 50QG1001, 50QG1101, 50\-QG1401, 50QG1402, 50QG1403, 50QG1404, 50QG1904, 50QG2101, 50QG2102, and 50QG2202, and the Centre for Information Services and High Performance Computing (ZIH) at the Technische Universit\"{a}t Dresden for generous allocations of computer time;
\item the Hungarian Academy of Sciences through the Lend\"{u}let Programme grants LP2014-17 and LP2018-7 and the Hungarian National Research, Development, and Innovation Office (NKFIH) through grant KKP-137523 (`SeismoLab');
\item the Science Foundation Ireland (SFI) through a Royal Society - SFI University Research Fellowship (M.~Fraser);
\item the Israel Ministry of Science and Technology through grant 3-18143 and the Tel Aviv University Center for Artificial Intelligence and Data Science (TAD) through a grant;
\item the Agenzia Spaziale Italiana (ASI) through contracts I/037/08/0, I/058/10/0, 2014-025-R.0, 2014-025-R.1.2015, and 2018-24-HH.0 to the Italian Istituto Nazionale di Astrofisica (INAF), contract 2014-049-R.0/1/2 to INAF for the Space Science Data Centre (SSDC, formerly known as the ASI Science Data Center, ASDC), contracts I/008/10/0, 2013/030/I.0, 2013-030-I.0.1-2015, and 2016-17-I.0 to the Aerospace Logistics Technology Engineering Company (ALTEC S.p.A.), INAF, and the Italian Ministry of Education, University, and Research (Ministero dell'Istruzione, dell'Universit\`{a} e della Ricerca) through the Premiale project `MIning The Cosmos Big Data and Innovative Italian Technology for Frontier Astrophysics and Cosmology' (MITiC);
\item the Netherlands Organisation for Scientific Research (NWO) through grant NWO-M-614.061.414, through a VICI grant (A.~Helmi), and through a Spinoza prize (A.~Helmi), and the Netherlands Research School for Astronomy (NOVA);
\item the Polish National Science Centre through HARMONIA grant 2018/30/M/ST9/00311 and DAINA grant 2017/27/L/ST9/03221 and the Ministry of Science and Higher Education (MNiSW) through grant DIR/WK/2018/12;
\item the Portuguese Funda\c{c}\~{a}o para a Ci\^{e}ncia e a Tecnologia (FCT) through national funds, grants SFRH/\-BD/128840/2017 and PTDC/FIS-AST/30389/2017, and work contract DL 57/2016/CP1364/CT0006, the Fundo Europeu de Desenvolvimento Regional (FEDER) through grant POCI-01-0145-FEDER-030389 and its Programa Operacional Competitividade e Internacionaliza\c{c}\~{a}o (COMPETE2020) through grants UIDB/04434/2020 and UIDP/04434/2020, and the Strategic Programme UIDB/\-00099/2020 for the Centro de Astrof\'{\i}sica e Gravita\c{c}\~{a}o (CENTRA);  
\item the Slovenian Research Agency through grant P1-0188;
\item the Spanish Ministry of Economy (MINECO/FEDER, UE), the Spanish Ministry of Science and Innovation (MICIN), the Spanish Ministry of Education, Culture, and Sports, and the Spanish Government through grants BES-2016-078499, BES-2017-083126, BES-C-2017-0085, ESP2016-80079-C2-1-R, ESP2016-80079-C2-2-R, FPU16/03827, PDC2021-121059-C22, RTI2018-095076-B-C22, and TIN2015-65316-P (`Computaci\'{o}n de Altas Prestaciones VII'), the Juan de la Cierva Incorporaci\'{o}n Programme (FJCI-2015-2671 and IJC2019-04862-I for F.~Anders), the Severo Ochoa Centre of Excellence Programme (SEV2015-0493), and MICIN/AEI/10.13039/501100011033 (and the European Union through European Regional Development Fund `A way of making Europe') through grant RTI2018-095076-B-C21, the Institute of Cosmos Sciences University of Barcelona (ICCUB, Unidad de Excelencia `Mar\'{\i}a de Maeztu’) through grant CEX2019-000918-M, the University of Barcelona's official doctoral programme for the development of an R+D+i project through an Ajuts de Personal Investigador en Formaci\'{o} (APIF) grant, the Spanish Virtual Observatory through project AyA2017-84089, the Galician Regional Government, Xunta de Galicia, through grants ED431B-2021/36, ED481A-2019/155, and ED481A-2021/296, the Centro de Investigaci\'{o}n en Tecnolog\'{\i}as de la Informaci\'{o}n y las Comunicaciones (CITIC), funded by the Xunta de Galicia and the European Union (European Regional Development Fund -- Galicia 2014-2020 Programme), through grant ED431G-2019/01, the Red Espa\~{n}ola de Supercomputaci\'{o}n (RES) computer resources at MareNostrum, the Barcelona Supercomputing Centre - Centro Nacional de Supercomputaci\'{o}n (BSC-CNS) through activities AECT-2017-2-0002, AECT-2017-3-0006, AECT-2018-1-0017, AECT-2018-2-0013, AECT-2018-3-0011, AECT-2019-1-0010, AECT-2019-2-0014, AECT-2019-3-0003, AECT-2020-1-0004, and DATA-2020-1-0010, the Departament d'Innovaci\'{o}, Universitats i Empresa de la Generalitat de Catalunya through grant 2014-SGR-1051 for project `Models de Programaci\'{o} i Entorns d'Execuci\'{o} Parallels' (MPEXPAR), and Ramon y Cajal Fellowship RYC2018-025968-I funded by MICIN/AEI/10.13039/501100011033 and the European Science Foundation (`Investing in your future');
\item the Swedish National Space Agency (SNSA/Rymdstyrelsen);
\item the Swiss State Secretariat for Education, Research, and Innovation through the Swiss Activit\'{e}s Nationales Compl\'{e}mentaires and the Swiss National Science Foundation through an Eccellenza Professorial Fellowship (award PCEFP2\_194638 for R.~Anderson);
\item the United Kingdom Particle Physics and Astronomy Research Council (PPARC), the United Kingdom Science and Technology Facilities Council (STFC), and the United Kingdom Space Agency (UKSA) through the following grants to the University of Bristol, the University of Cambridge, the University of Edinburgh, the University of Leicester, the Mullard Space Sciences Laboratory of University College London, and the United Kingdom Rutherford Appleton Laboratory (RAL): PP/D006511/1, PP/D006546/1, PP/D006570/1, ST/I000852/1, ST/J005045/1, ST/K00056X/1, ST/\-K000209/1, ST/K000756/1, ST/L006561/1, ST/N000595/1, ST/N000641/1, ST/N000978/1, ST/\-N001117/1, ST/S000089/1, ST/S000976/1, ST/S000984/1, ST/S001123/1, ST/S001948/1, ST/\-S001980/1, ST/S002103/1, ST/V000969/1, ST/W002469/1, ST/W002493/1, ST/W002671/1, ST/W002809/1, and EP/V520342/1.
\end{itemize}

The GBOT programme  uses observations collected at (i) the European Organisation for Astronomical Research in the Southern Hemisphere (ESO) with the VLT Survey Telescope (VST), under ESO programmes
092.B-0165,
093.B-0236,
094.B-0181,
095.B-0046,
096.B-0162,
097.B-0304,
098.B-0030,
099.B-0034,
0100.B-0131,
0101.B-0156,
0102.B-0174, and
0103.B-0165;
%
%
and (ii) the Liverpool Telescope, which is operated on the island of La Palma by Liverpool John Moores University in the Spanish Observatorio del Roque de los Muchachos of the Instituto de Astrof\'{\i}sica de Canarias with financial support from the United Kingdom Science and Technology Facilities Council, and (iii) telescopes of the Las Cumbres Observatory Global Telescope Network.

\end{appendix}

\end{document}